\documentclass[english,onecolumn]{IEEEtran}
\pdfoutput=1
\expandafter\let\csname equation*\endcsname\relax
\expandafter\let\csname endequation*\endcsname\relax
\usepackage{amsmath,amsfonts,amssymb}
\usepackage{amsthm}
\usepackage{graphicx}
\usepackage{algorithm}
\usepackage{algpseudocode}
\usepackage{color}
\usepackage{multirow}
\usepackage{tabularx}
\usepackage{array}
\usepackage{placeins}
\usepackage{subcaption}
\usepackage{comment}
\usepackage[style=ieee, backend=bibtex, mincitenames=1, maxcitenames=1, maxbibnames=99]{biblatex}
\usepackage{geometry}
\usepackage{listings}
\usepackage{color}
\usepackage{upgreek}
\usepackage[percent]{overpic}
\usepackage[switch]{lineno}

\bibliography{PGOSCTV}

\DeclareMathOperator*{\argmin}{argmin\,}

\DeclareMathOperator{\proj}{\mathcal{P}}
\DeclareMathOperator{\prox}{prox}

\DeclareMathOperator{\st}{such\,that\,}

\DeclareMathOperator{\diag}{diag}

\newcounter{pcounter}
\newenvironment{pequation}{\addtocounter{equation}{-1} 
	\refstepcounter{pcounter} 
	 
	\begin{equation}}
{\end{equation}\ignorespacesafterend}

\newcolumntype{C}{>{\centering\arraybackslash}X}

\begin{document}
\title{Technical Note: Proximal Ordered Subsets Algorithms for TV Constrained Optimization in  CT Image Reconstruction}%
\author{Sean Rose$^1$, Martin S. Andersen$^2$, Emil Y. Sidky$^1$, and Xiaochuan Pan$^1$
\thanks{
$^1$The University of Chicago,
Department of Radiology MC-2026,
5841 S. Maryland Avenue, Chicago IL, 60637}
\thanks{
$^2$Technical University of Denmark,
Department of Applied Mathematics and Computer Science,
Lyngby, Denmark}
}
\maketitle

\begin{abstract}
This article is intended to supplement our recent paper [\fullcite{Rose2015}] in which ordered subsets methods were
employed to perform total-variation (TV) constrained data-discrepancy minimization for image reconstruction in
X-ray computed tomography (CT). Here we provide details regarding implementation of the ordered subsets algorithms
and suggestions for selection of algorithm parameters. Detailed pseudo-code is included for every algorithm implemented
in the original manuscript.

\end{abstract}
%%%%%%%%%%%%%%%%%%%%%%%%%%%%%%%%%%%%%%%%%%%%%%%%%%%%%%%%%%%%
%%%%%%%%%%%%%%%%%%%%%%%%%%%%%%%%%%%%%%%%%%%%%%%%%%%%%%%%%%%%
%%%%%%%%%%%%%%%%%%%%%%%%%%%%%%%%%%%%%%%%%%%%%%%%%%%%%%%%%%%%

\section{Introduction}	
In our recent paper on the noise properties of images reconstructed using total-variation (TV) constrained data-discrepancy minimization \cite{Rose2015},
ordered subsets methods --- also referred to as incremental, row-action, and batch methods --- were employed to perform constrained optimization. This
was motivated in part by recent work in the optimization community in which the convergence properties of a general class of 
proximal-gradient ordered subsets algorithms were investigated \cite{Bertsekas2011, Andersen2014}. Here, detailed instructions and pseudo-code are provided for implementing ordered subsets algorithms for total-variation constrained
weighted least squares (TVC-WLSQ) and Poisson likelihood (TVC-PL) optimization in computed tomographic (CT) image reconstruction.  In section \ref{sec:IncAlg}
the design of the ordered subsets algorithms is outlined following the framework developed in \cite{Bertsekas2011}.  In section \ref{sec:TVProj} it is demonstrated how one 
can perform projection onto a TV-ball using a first order primal dual algorithm proposed by Chambolle and Pock \cite{Chambolle2010}.  In section \ref{sec:FullPseudo}
 full pseudo-code is provided for each algorithm with some additional notes and recommendations for implementation.

%%%%%%%%%%%%%%%%%%%%%%%%%%%%%%%%%%%%%%%%%%%%%%%%%%%%%%%%%%%%
%%%%%%%%%%%%%%%%%%%%%%%%%%%%%%%%%%%%%%%%%%%%%%%%%%%%%%%%%%%%
%%%%%%%%%%%%%%%%%%%%%%%%%%%%%%%%%%%%%%%%%%%%%%%%%%%%%%%%%%%%
\section{Algorithm Design}
\label{sec:IncAlg}
The proximal gradient ordered subsets framework applies to optimization problems in which the objective function can be separated
into a sum of component functions as follows
\begin{pequation}
	\min_{x \in X} \sum_{i=1}^S \left(g_i(x) + h_i(x) \right)
\end{pequation}
where $g_i, h_i: \Re^n \mapsto \Re$ are convex functions and $X$ is a convex set. Ordered subsets algorithms operate on individual component
functions one at a time instead of utilizing the entire objective function for every update. The ordered subsets algorithm used in our recent paper, and originally proposed
in \cite{Bertsekas2011}, has the form 
\begin{align}
	\label{eqn:update}
	&q^k = \prox_{t_k g_{i_k}}(x^k)\\ 
	&x^{k+1} = \proj(q^k - t_k \tilde{\nabla}h_{i_k}(q^k); X) \notag
\end{align}
where $\proj(\cdot; X)$ represents projection onto $X$, $\tilde{\nabla} h(q)$ represents a subgradient of $h$ at $q$, and
the proximal operator is defined as 
\begin{align*}
	\prox_f(x) = \argmin_u \left\{f(u) + \frac{1}{2} \|u-x\|_2^2 \right\}
\end{align*}
Here and throughout this document superscripts on vector quantities denote iterates of an algorithm, while superscripts attached to scalars denote raising the scalar to the given power.
The value of $i_k$ can be chosen in a cylic fashion ($i_k = k \mod S$) or in a randomized manner.

In this section, we derive ordered subsets algorithms for the TVC-WLSQ and TVC-PL reconstruction optimization problems using the update step in \ref{eqn:update}.

\subsection{TVC-WLSQ}
The TVC-WLSQ reconstruction optimization problem takes the form 
\begin{pequation}
	\begin{aligned}
		&\min_x \frac{1}{2} (Ax - b)^T \diag(w) (Ax-b)\\
		&\st \mathrm{TV}(x) \leq \gamma
	\end{aligned}
\end{pequation}
where $A \in \Re^{m\times n}$ is the system matrix representing the forward model, $b \in \Re^m$ is the measured sinogram data,
$x \in \Re^n$ is an image estimate, and $w \in \Re_{++}^n$ is a strictly positive weighting vector. An ordered subsets
algorithm can be derived by defining  
\begin{align*}
	&g_i(x) = \frac{1}{2}w_i(a_i^Tx-b_i)^2 \text{ for } i=1,\ldots,m \\
	&h_i(x) = 0 \text{ for } i=1,\ldots,m+1\\
\end{align*}
and
\begin{align*}
	g_{m+1}(x) = \delta(x; B_{\mathrm{TV}}(\gamma))
\end{align*}
where 
\begin{align*}
	\delta(x; X) = \begin{cases} 0 &\text{ if } x \in X\\
						\infty &\text{ else }
						\end{cases}
\end{align*}
and $B_{\mathrm{TV}}(\gamma) = \{x: \mathrm{TV}(x) \leq \gamma\}$ is the TV ball of radius $\gamma$.  The $i^{th}$ row of the system matrix $A$ has been denoted $a_i^T$. 
 Note that in the definition of $g_{m+1}$ the constraint 
has been imposed using an indicator function and $X$ can therefore be taken to be $\Re^n$.

It can be shown that \cite{Rose2015}
\begin{align*}
	&\prox_{t_k g_{i_k} }(x) = x - \frac{a_{i_k}^T x -b_{i_k}}{(w_{i_k} t_k)^{-1} + \|a_{i_k}\|_2^2}a_{i_k} \text{ for } i_k=1,\ldots,m \\
	&\prox_{t_k g_{m+1}}(x) = \proj(x,B_{\mathrm{TV}}(\gamma))
\end{align*}
This yields the following algorithm
\FloatBarrier
\begin{algorithm}[h!]                      % enter the algorithm environment
\caption{Ordered Subsets TVC-WLSQ }          % give the algorithm a caption
\label{alg:ITVCWLS}                           % and a label for \ref{} commands later in the document
\begin{algorithmic}[1]
\State Initialize $x_0$ to zero
\For{$k=0,\ldots, K-1$}
\State	$p^0 \gets x^k$
\For{$i=0,\ldots,m-1$}
\State	$p^{i+1} \gets p^i - \frac{a_i^T p^i - b_i}{\|a_i\|_2^2 + (t_k w_{i})^{-1}} a_i$
\EndFor
\State $x^{k+1} \gets \proj(p^m,B_{\mathrm{TV}}(\gamma))$
\EndFor
\end{algorithmic}
\end{algorithm}
\FloatBarrier
The sub-iterations for $p$ are similar in nature to the algebraic reconstruction technique (ART) \cite{KakSlaney}, involving a loop over the data with an image update based 
upon each ray.   Note that by imposing the TV constraint as an indicator function one obtains an algorithm which
involves projection onto the constraint set only once per loop over the data. 
The projection onto the TV-ball must be done using a numerical method.  In section \ref{sec:TVProj},
it is shown how to perform this operation with the primal-dual algorithm of Chambolle and Pock (CP algorithm) \cite{Chambolle2010}.

The choice of step-size $t_k$ can have a significant effect on convergence rate.  To ensure convergence,
a diminishing step-size is required.  We have found the following
step-size rule to be useful in practice
\begin{align*}
	&c_k = \left\lfloor \frac{k}{r} \right\rfloor + 1\\
	&t_k = \frac{1}{c_k}
\end{align*}
where $r$ represents the number of steps for which we hold the step-size constant and $\left\lfloor \cdot \right\rfloor$ denotes the floor operation.  In our studies, $r$ is often
taken to be $20$, but we suggest experimentation to determine a reasonable value for any given scenario.

\subsection{TVC-PL}
The reconstruction optimization problem for TVC-PL is given by
\begin{pequation}
	\begin{aligned}
		&\min_x \sum_{i=1}^m \left( y_i a_i^T x + N_0 \exp(-a_i^Tx) \right)\\
		&\st \mathrm{TV}(x) \leq \gamma
	\end{aligned}
\end{pequation}
where $y \in \Re^m$ is the measured transmission data and $N_0$ is the number of incident photons per ray. An incremental algorithm is derived from the update step in \ref{eqn:update} using the following definitions
\begin{align*}
	&g_i(x) = y_i a_i^Tx + N_0 \exp(-a_i^Tx) \text{ for } i=1,\ldots,m \\
	&h_i(x) = 0 \text{ for } i=1,\ldots,m+1\\
\end{align*}
and
\begin{align*}
	g_{m+1}(x) = \delta(x; B_{\mathrm{TV}}(\gamma))
\end{align*}
It can be shown that \cite{Rose2015}
\begin{align*}
	\prox_{t_k g_{i_k} }(x^k) = x^k  + t_k(N_0 \exp(-c_k^*) - y_{i_k})a_{i_k} \text{ for } i_k=1,\ldots,m
\end{align*}
where $c_k^*$ is found by implicit solution of
\begin{align*}
	c_k^* = a_{i_k}^T x  + t_k \|a_{i_k}\|_2^2(N_0 \exp(-c_k^*) -y_{i_k})
\end{align*}
Methods for calculating $c_k^*$ are provided in \cite{Rose2015}. The proximal operation for $g_{m+1}$ is the same as for TVC-WLSQ.

Plugging the derived proximal updates into \ref{eqn:update} yields the following algorithm.

\FloatBarrier
\begin{algorithm}[h!]                      % enter the algorithm environment
\caption{Ordered Subsets TVC-PL }          % give the algorithm a caption
\label{alg:ITVCPL}                           % and a label for \ref{} commands later in the document
\begin{algorithmic}[1]
\State Initialize $x_0$ to zero
\For{$k=0,\ldots, K-1$}
\State	$p^0 \gets x^k$
\For{$i=0,\ldots,m-1$}
\State 	Solve $c_i^* = a_{i}^T p^i  + t_k \|a_{i}\|_2^2(N_0 \exp(-c_i^*) -y_{i})$ for $c_i^*$
\State	$p^{i+1} \gets p^i  + t_k(N_0 \exp(-c_i^*) - y_{i})a_{i}$
\EndFor
\State $x^{k+1} \gets \proj(p^m,B_{\mathrm{TV}}(\gamma))$
\EndFor
\end{algorithmic}
\end{algorithm}
\FloatBarrier

%%%%%%%%%%%%%%%%%%%%%%%%%%%%%%%%%%%%%%%%%%%%%%%%%%%%%%%%%%%%
%%%%%%%%%%%%%%%%%%%%%%%%%%%%%%%%%%%%%%%%%%%%%%%%%%%%%%%%%%%%
%%%%%%%%%%%%%%%%%%%%%%%%%%%%%%%%%%%%%%%%%%%%%%%%%%%%%%%%%%%%
\section{Chambolle-Pock Algorithm for projection onto TV ball: Derivation and Pseudo-code}
\label{sec:TVProj}
Here the derivation and pseudo-code of an instance of the CP algorithm are presented
for projection of an image onto a TV-ball of radius $\gamma$.  
The derivation follows the framework and notation presented by Sidky et al. \cite{Sidky2012} in which the algorithm
was applied to a variety of optimization problems for CT image reconstruction.

The Chambolle-Pock algorithm is used to solve convex optimization problems written in the form
\begin{align*}
	\min_s F(Ks) + G(s)
\end{align*}
Here the algorithm is employed to solve the problem of projecting onto $B_{\mathrm{TV}}(\gamma)$, which can be expressed as
\begin{gather*}
	\min_s \frac{1}{2}\|s -x \|_2^2\\
	\st \|h(D_1 s, D_2 s)\|_1 \leq \gamma
\end{gather*}
where $h: \Re^{2n} \mapsto \Re^n$ is defined by 
\begin{align*}
	h_i(y,z) = \sqrt{y_i^2 + z_i^2}
\end{align*}
The operators $D_1: \Re^n \mapsto \Re^n$ and $D_2: \Re^n \mapsto \Re^n$ are finite difference operators which calculate
approximations of the $x$ and $y$ components of the spatial gradients of the image, respectively.

For the purposes of applying the algorithm, define
\begin{align*}
	&F(y,z) =  \delta(h(y,z);B_1( \gamma))\\
	&G(s) = \frac{1}{2}\|s-x\|_2^2\\
	&K = \begin{pmatrix}D_1\\ D_2\end{pmatrix}
\end{align*}
where  $B_1(\gamma) = \{s: \|s\|_1 \leq \gamma\}$.

In this section, pseudo-code is first presented for the instance of the CP algorithm these 
definitions yield. A detailed derivation of the proximal mappings used in the pseudo-code follows.

\subsection{TV Projection Pseudo-code}
The Chambolle Pock update for projection onto the TV ball takes the form
\FloatBarrier
\begin{algorithm}[h!]                      % enter the algorithm environment
\caption{Projection onto TV ball}          % give the algorithm a caption
\label{alg:TVProj}                           % and a label for \ref{} commands later in the document
\begin{algorithmic}[1]
\State $L \gets \left\| \begin{pmatrix} D_1 \\ D_2 \end{pmatrix} \right\|_2$; \, $\tau \gets 1/L$; \, $\sigma \gets 1/L$; \, $\theta \gets 1$; \, $k \gets0$
\State Initialize $s_0$, $y_0$, and $z_0$
\State $\bar{s} \gets s_0$
\For{$k = 0, \ldots, K-1$}
	\State $y^k \gets y^k + \sigma  D_1\bar{s}^k$ 
	\State $z^k \gets z^k + \sigma  D_2\bar{s}^k$
	\vspace{0.1cm}
	\State $v \gets \frac{\proj(h(y^k,z^k)/\sigma;B_1(\gamma))}{h(y^k,z^k)}$
	\vspace{0.1cm}
	\State $\begin{pmatrix} y^{k+1}\\z^{k+1} \end{pmatrix}\gets \begin{pmatrix} y^k \\ z^k \end{pmatrix}
	 - \sigma \begin{pmatrix} \diag(v) & 0 \\ 0 & \diag(v) \end{pmatrix}
	 \vspace{0.1cm}
	\begin{pmatrix} y^k \\ z^k \end{pmatrix}$
	\State $ s^{k+1} \gets s^k - \tau  (D_1^T, D_2^T)\begin{pmatrix} y^{k+1}\\z^{k+1}\end{pmatrix}$
	\State $ s^{k+1} \gets \frac{s^{k+1}/\tau +  b}{1 + 1/\tau}$
	\State $\bar{s}^{k+1} = s^{k+1} + \theta (s^{k+1} -s^k)$
\EndFor
\end{algorithmic}
\end{algorithm}
\FloatBarrier
Calculating the $\ell_2$ norm of a matrix --- needed to calculate $L$ --- can be done with the power method \cite{Sidky2012}. 
In order to implement the projection algorithm, one must also be able to perform projection onto the $\ell_1$ ball.  This can be done using
the method of Duchi et al. \cite{Duchi2008}, which is presented as algorithm 2 in another of our recent papers \cite{Sidky2014}.  In practice, 
the TV-projection algorithm is only run for approximately 10 iterations, yielding an inexact projection onto the TV-ball.  It is important
when running the algorithm to monitor the TV of the individual iterates to make sure the algorithm yields a result which is 
inside or close to the TV-ball.

\subsection{Proximal Mapping of G}
Here the proximal mapping of G
\begin{gather*}
	\prox_{\tau G}(s) = \argmin_u \left\{ \frac{1}{2} \|u-x\|_2^2 + \frac{\|u-s\|_2^2}{2 \tau} \right\}
\end{gather*}
is derived.
The first order optimality conditions yield
\begin{gather*}
	(u^*-x)+\frac{1}{\tau}(u^*-s) = 0 \\
	\implies \left(1+\frac{1}{\tau}\right)u^* = x + s/\tau\\
	\implies \prox_{\tau G}(s) = \frac{ x + s/\tau}{1+1/\tau}
\end{gather*}

\subsection{Proximal Mapping of $F^*$}
To find the proximal mapping of $F^*$, one can employ the Moreau identity
\begin{align*}
	\prox\limits_{\sigma F^*}(y,z) = \begin{pmatrix} y\\z\end{pmatrix} - \sigma \prox\limits_{F/\sigma}(y/\sigma,z/\sigma)
\end{align*}
It is then necessary to evaluate
\begin{align*}
	\prox\limits_{F}(y,z) = \argmin_{\substack{u, w \\ h(u,w) \in B_1( \gamma)}}\{\|u-y\|_2^2 + \|w-z\|_2^2\}
\end{align*}
To do so, note that the minimization can be separated component-wise.  It can then be seen
\begin{align*}
	(u_i-y_i)^2 + (w_i - z_i)^2 &= (u_i^2 + w_i^2) + (y_i^2+z_i^2) - 2(u_i y_i + w_i z_i)\\
	&\geq (u_i^2 + w_i^2) + (y_i^2 +z_i^2) - 2 \|(u_i,w_i)^T\|\|(y_i,z_i)^T\|\\
	&=(h_i(u,w) - h_i(y,z))^2
\end{align*}
where the second line follows from the Cauchy-Schwarz inequality.  Equality holds throughout if
\begin{align*}
	(u_i,w_i)^T = c_i (y_i,z_i)^T, \, i = 1,\ldots, n; \, c_i \geq 0 
\end{align*}
One can therefore evaluate the proximal mapping by first projecting $h(y,z)$ onto the
$\ell_1$ ball of radius $\gamma $ yielding a vector $g^* \in \Re^n$.  One must then choose
$(u,w)$ such that $h(u,w) = g^*$ and $(u_i,w_i)^T = c_i (y_i,z_i)^T$.  This can be done by 
scaling each vector $(y_i,z_i)$ by the amount $c_i  = g_i^*/\|(y_i,z_i)^T\|_2$ if $\|(y_i,z_i)^T\|_2 > 0$, 
and $c_i = 0$ otherwise.  It follows that
\begin{align*}
	\prox_{F}(y,z) &= \begin{pmatrix} \diag\left[\frac{\proj(h(y,z);B_1(\gamma))}{h(y,z)}\right] & 0 \\
							0 & \diag\left[\frac{\proj(h(y,z);B_1(\gamma))}{h(y,z)}\right]  \end{pmatrix}
							\begin{pmatrix} y \\ z \end{pmatrix}
\end{align*}
where division of vectors is to be interpreted 
element-wise.  Also note the slight abuse of notation in that the $i^{th}$ element of 
$\frac{\proj(h(y,z);B_1(\gamma))}{h(y,z)}$ should be $0$ if $h_i(x,y)$ is $0$. By substituting variables, one finds
\begin{gather*}
	\prox_{F/\sigma}(y/\sigma,z/\sigma) = \begin{pmatrix} \diag\left[\frac{\proj(h(y/\sigma,z/\sigma);B_1( \gamma))}{h(y,z)}\right] & 0 \\
																0 & \diag\left[\frac{\proj(h(y/\sigma,z/\sigma);B_1( \gamma))}{h(y,z)}\right] \end{pmatrix}
	\begin{pmatrix} y \\ z \end{pmatrix}
\end{gather*}
Plugging into the Moreau identity yields
\begin{gather*}
	\prox_{\sigma F^*}(y,z) = \begin{pmatrix} y \\ z \end{pmatrix} - \sigma \begin{pmatrix} \diag\left[\frac{\proj(h(y/\sigma,z/\sigma);B_1( \gamma))}{h(y,z)}\right] & 0\\
	 0 & \diag\left[\frac{\proj(h(y/\sigma,z/\sigma);B_1( \gamma))}{h(y,z)}\right] \end{pmatrix}
	\begin{pmatrix} y \\ z \end{pmatrix}
\end{gather*}

%%%%%%%%%%%%%%%%%%%%%%%%%%%%%%%%%%%%%%%%%%%%%%%%%%%%%%%%%%%%
%%%%%%%%%%%%%%%%%%%%%%%%%%%%%%%%%%%%%%%%%%%%%%%%%%%%%%%%%%%%
%%%%%%%%%%%%%%%%%%%%%%%%%%%%%%%%%%%%%%%%%%%%%%%%%%%%%%%%%%%%
\section{The Full Algorithm}
\label{sec:FullPseudo}
Here the full pseudo-code for the ordered subsets TVC-WLSQ and TVC-PL is presented.
As noted previously, in our experience the parameter $J$, the number of iterations of the TV-projection
algorithm,  can typically be taken to be 10 and the step size chosen according according to the rule 
\begin{align*}
	&c_k = \left\lfloor \frac{k}{r} \right\rfloor + 1\\
	&t_k = \frac{1}{c_k}
\end{align*}
with $r=20$.

\FloatBarrier
\begin{algorithm}[h!]                      % enter the algorithm environment
\caption{Ordered Subsets TVC-WLSQ}               % give the algorithm a caption
\label{alg:FullITVCWLS}                           % and a label for \ref{} commands later in the document
\begin{algorithmic}[1]
\State $L \gets \left \| \begin{pmatrix} D_1 \\ D_2 \end{pmatrix}\right\|_2$; \, $\tau \gets 1/L$; \, $\sigma \gets 1/L$; \, $\theta \gets 1$; \, $k \gets0$

\State Initialize $x^0$, $s^0$, $y^0$, and $z^0$ to zero \Comment{Begin Iterating}
\For{$k=0,\ldots, K-1$}
\State Update step size $t_k$
\State	$p^0 \gets  x^k$	\For{$i=0,\ldots,m-1$} \Comment{Run Incremental Update}
	\State	$p^{i+1} \gets p^i - \frac{a_i^T p^i - b_i}{\|a_i\|_2^2 + (t_k w_{i})^{-1}} a_i$
\EndFor
	\State $x^{k+1} \gets p^{m}$
	\If{$\mathrm{TV}(x^{k+1}) > \gamma$}	\Comment{Check TV}
		\State $\bar{s}^0 \gets s^0$
		\For{$j =0,\ldots, J-1$}	\Comment{Run Projection onto TV Ball}
			\State $y^j \gets y^j + \sigma  D_1\bar{s}^j$ 
			\State $z^j \gets z^j + \sigma  D_2\bar{s}^j$
			\vspace{0.1cm}
			\State $v \gets \frac{\proj(h(y^j/\sigma,z^j/\sigma);B_1(\gamma))}{h(y^j,z^j)}$
			\vspace{0.1cm}
			\State $\begin{pmatrix} y^{j+1}\\z^{j+1} \end{pmatrix}\gets \begin{pmatrix} y^j \\ z^j \end{pmatrix}
			 - \sigma \begin{pmatrix} \diag(v) & 0 \\ 0 &\diag(v) \end{pmatrix}
			\begin{pmatrix} y^j \\ z^j \end{pmatrix}$
			\State $ s^{j+1} \gets s^j - \tau  (D_1^T, D_2^T)\begin{pmatrix} y^{j+1}\\z^{j+1}\end{pmatrix}$
			\State $ s^{j+1} \gets \frac{s^{j+1}/\tau +  x^{k+1}}{1 + 1/\tau}$
			\State $\bar{s}^{j+1} = s^{j+1} + \theta (s^{j+1} -s^j)$
		\EndFor
		\State $x^{k+1} \gets s^{J}$
		\State $s^0 \gets s^{J}$	\Comment{Save Variables for Warm Start}
		\State $y^0 \gets y^{J}$
		\State $z^0 \gets z^{J}$
	\EndIf
	
\EndFor

\end{algorithmic}
\end{algorithm}

\FloatBarrier

As shown above,  a warm start is used for the projection algorithm, by which it is meant that the last iterates of the $s$, $y$, and $z$ variables
are used to initialize $s_0$, $y_0$, and $z_0$ the next time projection needs to be performed.
Since the CP algorithm for projection onto the TV ball is truncated at early iteration, it is imperative that one monitors the TV of the iterates of the projection
algorithm to ensure that it is performing correctly and that it outputs vectors which are close to or within the constraint set. The corresponding algorithm for
TVC-PL is identical except for line 7, which must be replaced by lines 5 and 6 in Algorithm \ref{alg:ITVCPL}.

%%%%%%%%%%%%%%%%%%%%%%%%%%%%%%%%%%%%%%%%%%%%%%%%%%%%%%%%%%%%
%%%%%%%%%%%%%%%%%%%%%%%%%%%%%%%%%%%%%%%%%%%%%%%%%%%%%%%%%%%%
%%%%%%%%%%%%%%%%%%%%%%%%%%%%%%%%%%%%%%%%%%%%%%%%%%%%%%%%%%%%
\printbibliography

\end{document}